\documentclass{article}
\usepackage{ijcai22}

\usepackage{amsmath,amsfonts,bm}

\def\eqref#1{equation~\ref{#1}}

\def\1{\bm{1}}

\newcommand{\beps}{\boldsymbol\epsilon}

\def\rvepsilon{{\boldsymbol{\epsilon}}}

\def\rvx{{\mathbf{x}}}

\def\vzero{{\bm{0}}}

\def\ve{{\bm{e}}}

\def\vx{{\bm{x}}}

\def\mF{{\bm{F}}}
\def\mG{{\bm{G}}}

\def\mI{{\bm{I}}}

\def\mBeta{{\bm{\beta}}}

\DeclareMathAlphabet{\mathsfit}{\encodingdefault}{\sfdefault}{m}{sl}
\SetMathAlphabet{\mathsfit}{bold}{\encodingdefault}{\sfdefault}{bx}{n}

\def\gL{{\mathcal{L}}}

\def\gN{{\mathcal{N}}}

\def\sR{{\mathbb{R}}}

\usepackage{times}
\usepackage{soul}
\usepackage{url}
\usepackage[hidelinks]{hyperref}
\usepackage[utf8]{inputenc}
\usepackage[small]{caption}
\usepackage{graphicx}
\usepackage{amsmath}
\usepackage{amsthm}
\usepackage{booktabs}
\usepackage{algorithm}
\usepackage{algorithmic}
\usepackage{multirow} 
\usepackage{amsmath} 
\usepackage{xcolor}
\usepackage{wrapfig}
\usepackage{amssymb}
\usepackage{subfigure}
\usepackage[misc,geometry]{ifsym}
\title{FastDiff: A Fast Conditional Diffusion Model for High-Quality Speech Synthesis}

\author{
Rongjie Huang$^1$\footnote{Work done during internship at Tencent AI Lab}\footnote{Equal contribution}\and
Max W. Y. Lam$^2$\footnotemark[2]\and
Jun Wang$^2$\and
Dan Su$^2$\and
Dong Yu$^3$\and
Yi Ren$^1$\and
Zhou Zhao$^1$\footnote{Corresponding author}\\
\affiliations
$^1$Zhejiang University \quad
$^2$Tencent AI Lab, China \quad
$^3$Tencent AI Lab, USA\\
\emails
\{rongjiehuang, rayeren, zhouzhao\}@zju.edu.cn,
\{maxwylam, joinerwang, dansu, dyu\}@tencent.com
}

\begin{document}

\maketitle
\begin{abstract}
Denoising diffusion probabilistic models (DDPMs) have recently achieved leading performances in many generative tasks. However, the inherited iterative sampling process costs hindered their applications to speech synthesis. This paper proposes FastDiff, a fast conditional diffusion model for high-quality speech synthesis. FastDiff employs a stack of time-aware location-variable convolutions of diverse receptive field patterns to efficiently model long-term time dependencies with adaptive conditions. A noise schedule predictor is also adopted to reduce the sampling steps without sacrificing the generation quality. Based on FastDiff, we design an end-to-end text-to-speech synthesizer, FastDiff-TTS, which generates high-fidelity speech waveforms without any intermediate feature (e.g., Mel-spectrogram). Our evaluation of FastDiff demonstrates the state-of-the-art results with higher-quality (MOS 4.28) speech samples. Also, FastDiff enables a sampling speed of 58x faster than real-time on a V100 GPU, making diffusion models practically applicable to speech synthesis deployment for the first time. We further show that FastDiff generalized well to the mel-spectrogram inversion of unseen speakers, and FastDiff-TTS outperformed other competing methods in end-to-end text-to-speech synthesis.\footnote{Audio samples are available at \url{https://FastDiff.github.io/}.}

\end{abstract}


\section{Introduction} \label{intro}

With the recent development of deep generative models, speech synthesis has seen an extraordinary progress. Among the conventional speech synthesis methods, WaveNets~\cite{oord2016wavenet} were demonstrated to generate high-fidelity audio samples in an autoregressive manner yet suffering from prohibitively expensive computational costs. In contrast, non-autoregressive approaches such as flow-based and GAN-based models~\cite{prenger2019waveglow,jang2021univnet,kong2020hifi,huang2021multi} were also proposed to generate speech audios with satisfactory speed. However, these models were still criticized for other problems, e.g., the limited sample quality or sample diversity \cite{xiao2021tackling}.

In speech synthesis, our goal is mainly two-fold:
\begin{itemize}
    \item High-quality: generating high-quality speech is a challenging problem especially when the sampling rate of an audio is high. It is vital to reconstruct details at different timescales for waveforms of highly variable patterns.
    \item Fast: high generation speed is essential when considering real-time speech synthesis. This poses a challenge for all high-quality neural synthesizers.
\end{itemize}

As a blossoming class of generative models, denoising diffusion probabilistic models (DDPMs)~\cite{ho2020denoising,song2020denoising,lam2022bddm,liu2022pseudo} has emerged to prove its capability to achieve leading performances in both image and audio syntheses~\cite{dhariwal2021diffusion,san2021noise,kong2020diffwave,chen2020wavegrad,lam2022bddm}. However, current development of DDPMs in speech synthesis was hampered by two major challenges:
\begin{itemize}
\item Different from other existing generative models, diffusion models are not trained to directly minimize the difference between the generated audio and the reference audio, but to de-noise a noisy sample given an optimal gradient. This in practice could lead to overly de-noised speech after a large number of sampling steps, in which natural voice characteristics including breathiness and vocal fold closure are removed.
\item While DDPMs inherently are gradient-based models, a guarantee of high sample quality typically comes at a cost of hundreds to thousands of de-noising steps. When reducing the sampling steps, an apparent degradation in quality due to perceivable background noise is observed.
\end{itemize}

In this work, we propose FastDiff, a fast conditional diffusion model for high-quality speech synthesis. To improve audio quality, FastDiff adopts a stack of time-aware location-variable convolutions of diverse receptive field patterns to efficiently model long-term time dependencies with adaptive conditions. To accelerate the inference procedure, FastDiff also includes a noise schedule predictor, which derives a short and effective noise schedule and significantly reduces the de-noising steps. Based on FastDiff, we also introduce an end-to-end phoneme-to-waveform synthesizer FastDiff-TTS, which simplifies the text-to-speech generation pipeline and does not require intermediate features or specialized loss functions to enjoy low inference latency.

Experimental results demonstrated that FastDiff achieved a higher MOS score than the best publicly available models and outperformed the strong WaveNet vocoder (MOS: 4.28 vs. 4.20). FastDiff further enjoys an effective sampling process and only needs 4 iterations to synthesize high-fidelity speech, 58x faster than real-time on a V100 GPU without engineered kernels. To the best of our knowledge, FastDiff is the first diffusion model with a sampling speed comparable to previous  for the first time applicable to interactive, real-world speech synthesis applications at a low computational cost. FastDiff-TTS successfully simplify the text-to-speech generation pipeline and outperform competing architectures.

  \section{Background: Denoising Diffusion Probabilistic Models}
  Denoising diffusion probabilistic models (DDPMs)~\cite{ho2020denoising,song2020denoising,lam2022bddm} are likelihood-based generative models that have recently succeeded to advance the state-of-the-art results in benchmark generative tasks \cite{dhariwal2021diffusion} and have proved its capability to produce high-quality samples. The basic idea of DDPMs is to train a gradient neural network for reversing a diffusion process. Given i.i.d. samples $\{\vx_{0} \in \mathbb{R}^{D}\}$ from an unknown data distribution $p_{data}(\vx_{0})$, DDPMs try to approximate $p_{data}(\vx_{0})$ by a marginal distribution $p_{\theta}(\vx_{0})=\int\cdots\int p(\vx_T)\prod_{t=1}^{T} p_{\theta}(\vx_{t-1}|\vx_{t}) d \vx_1 \ldots d\vx_T$.
  
  In data distribution as $q(\vx_{0})$, the diffusion process is defined by a fixed Markov chain from data $\vx_0$ to the latent variable $\vx_T$. 
For a small positive constant $\beta_t$, a small Gaussian noise is added from $\vx_{t}$ to the distribution of $\vx_{t-1}$ under the function of $q(\vx_t|\vx_{t-1})$. The whole process gradually converts data $\vx_0$ to whitened latents $\vx_T$ according to the fixed noise schedule $\beta_1,\cdots,\beta_T$. 
  The reverse process is a Markov chain from $\vx_T$ to $\vx_0$ parameterized by a shared $\theta$, which aims to recover samples from Gaussian noises though eliminating the Gaussian noise added in the diffusion process in each iteration.
  
  It has been demonstrated that diffusion probabilistic models~\cite{dhariwal2021diffusion,xiao2021tackling} can learn diverse data distribution in multiple domains, such as images and time series. While the main issue with the proposed neural diffusion process is that it requires up to thousands of iterative steps to reconstruct the target distribution during reverse sampling. In this work, we offer a fast conditional diffusion model to reduce reverse iterations and improve computational efficiency. 
\begin{figure*}
    \vspace{-2mm}
    \centering
    \includegraphics[width=0.95\textwidth,trim={1.0cm 0cm 1.0cm 0cm}]{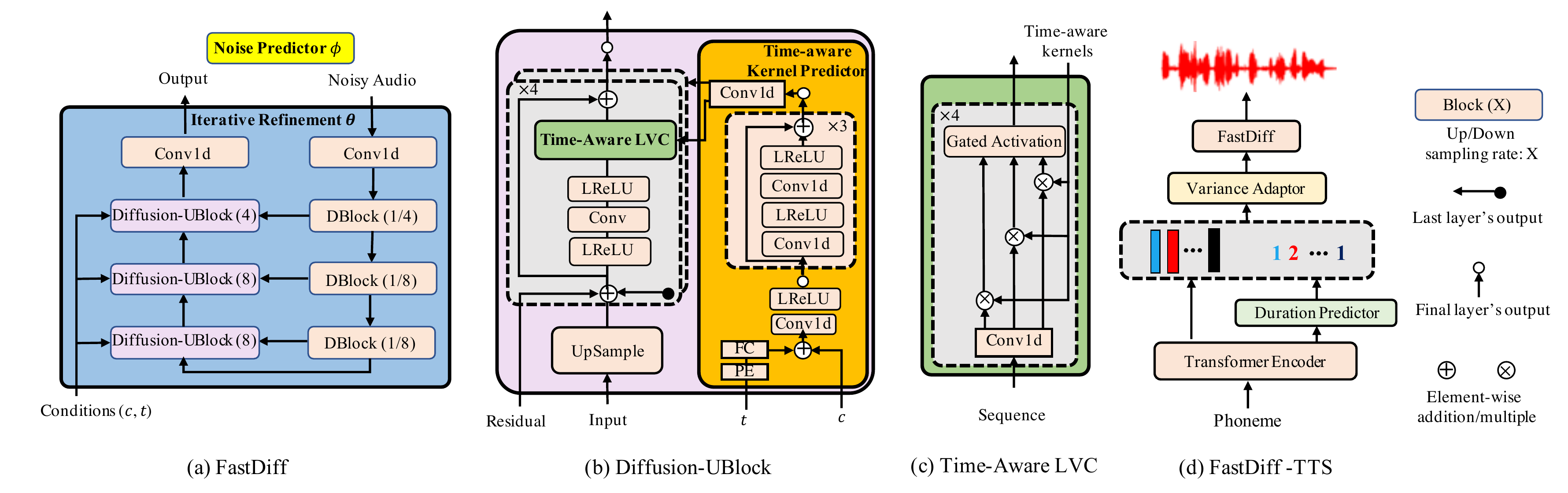}
    \vspace{-1mm}
   \caption{The overall architecture for FastDiff and FastDiff-TTS. The refinement model $\theta$ in FastDiff takes noisy audio $\vx_{t}$ as input and computes $\beps_{\theta}(\vx_{t}|c, t)$ conditioned on diffusion time index $t$ and Mel-spectrogram $c$. We use LReLU to denote the leaky rectified linear unit, LVC to denote the location-variable convolution, FC to denote the fully-connected layer, and PE to denote the positional encoding operation.} 
    \label{fig:arch}
  \end{figure*}

\section{FastDiff}
This section presents our proposed FastDiff, a fast conditional diffusion model for high-quality speech synthesis. We first describe the motivation of the design in FastDiff. Secondly, we introduce the iterative refinement model $\theta$ for high-quality speech synthesis and the noise predictor $\phi$ for accelerated sampling. Furthermore, we describe the training and inference procedures in detail. At last, we extend FastDiff to FastDiff-TTS for fully end-to-end text-to-speech syntheses.

\subsection{Motivation}
While denoising diffusion probabilistic models have shown high potential in synthesizing high-quality speech samples \cite{chen2020wavegrad,kong2020diffwave,liu2021diffsinger}, several challenges remain for industrial deployment: 1) Different from the traditional generative models, diffusion models catch dynamic dependencies from noisy audio instead of clean ones, which introduce more variation information (i.e, noise levels) in addition to the spectrogram fluctuation. 2) With limited receptive field patterns, a distinct degradation could emerge when reducing the reverse iterations, making diffusion models difficult to get accelerated. As a result, hundred or thousand orders of iterations prevent existing diffusion models from real-world deployment.

In FastDiff, we propose two key components to complement the above issues: 1) FastDiff adopts a time-aware location-variable convolution to catch the details of noisy samples at dynamic dependencies. The convolution operations are conditioned on dynamic variations in speech including diffusion steps and spectrogram fluctuations, equipping the model with diverse receptive field patterns and promoting the robustness of diffusion models during reverse acceleration. 2) To accelerate the inference procedure, FastDiff adopts a noise schedule predictor to reduce the number of reverse iterations, frees diffusion models from hundreds or thousands of refinement iterations. This makes FastDiff for the first time applicable to interactive, real-world applications at a low computational cost.


\subsection{Time-Aware Location-Variable Convolution}


In comparison with traditional convolution networks, location-variable convolution~\cite{zeng2021lvcnet} shows efficiency in modeling the long-term dependency of audio and gets neural network free from a significant number of dilated convolution layers. Inspired by this, we introduce the Time-Aware Location-Variable Convolution, which is sensitive to time steps in diffusion probabilistic models. At time step $t$, we follow~\cite{vaswani2017attention} to embed the step index into an 128-dimensional positional encoding (PE) vector $\ve_t$:
\begin{align*}
\ve_t=&\left[\sin \left(10^{\frac{0 \times 4}{63}} t\right), \cdots, \sin \left(10^{\frac{63 \times 4}{63}} t\right),\right.\\
&\,\,\left.\cos \left(10^{\frac{0 \times 4}{63}} t\right), \cdots, \cos \left(10^{\frac{63 \times 4}{63}} t\right)\right],
\end{align*}

In time-aware location-variable convolution, FastDiff requires multiple predicted variation-sensitive kernels to perform convolutional operations on the associated intervals of input sequence. These kernels should be time-aware and sensitive to variations of noisy audio including diffusion steps and acoustic features (i.e., Mel-spectrogram). Therefore, we propose a time-aware location-variable convolution (LVC) module, which is coupled with a kernel predictor as shown in Figure~\ref{fig:arch}(b) and Figure~\ref{fig:arch}(c). We describe the overall calculations below.

For the $q$-th time-aware LVC layer, we split the input $\vx_t\in\sR^{D}$ using a $M$-length window with $3^q$ dilations to produce $K$ segments with each $\vx_{t}^{k}\in\sR^M$:
  \begin{equation}
    \{\vx_{t}^{1}, \ldots, \vx_{t}^{K}\}=\operatorname{split}(\vx_t;M, q)
\end{equation}

Next, we perform convolutional operations on the associated intervals of input sequence using the kernels generated by a kernel predictor $\alpha$:
\begin{align}
    \{ \mF_{t}, \mG_{t}\} &= \alpha (t, c) \\ 
    \boldsymbol{z}_{t}^{k} &=\tanh (\boldsymbol{F}_{t} *  \boldsymbol{x}_{t}^{k}) \odot \sigma(\boldsymbol{G}_{t} *  \boldsymbol{x}_{t}^{k})\\
    \boldsymbol{z}_{t}&=\operatorname{concat}(\{\boldsymbol{z}_{t}^{1}, \ldots, \boldsymbol{z}_{t}^{K}\}),
\end{align}
where $\boldsymbol{F}_{t}, \boldsymbol{G}_{t}$ denote the filter and the gate kernels for $\vx_t^i$, respectively, $*$ denotes the 1d convolution, $\odot$ denotes the element-wise product and $\operatorname{concat}(\cdot)$ denotes the concatenation between vectors. Since the time-aware kernels are adaptive to the noise-level and dependent to the acoustic features, FastDiff is capable of precisely estimating de-noising gradient with a superior speed given a noisy signal input.

\subsection{Accelerated Sampling}

\subsubsection{Noise Predictor}
To avoid sampling with hundreds to thousands steps, FastDiff adopts the noise scheduling algorithm in the bilateral denoising diffusion models (BDDMs)~\cite{lam2022bddm} to predict a sampling noise schedule much shorter than the noise schedule used in training. This scheduling method has been revealed to be superior than other sampling acceleration methods, e.g.,  the grid search algorithm in WaveGrad~\cite{chen2020wavegrad} and the fast sampling algorithm in DiffWave~\cite{kong2020diffwave}. The noise predictor iteratively derives a continuous noise schedule $\hat{\mBeta}\in\sR^{T_m}$. We attach the learning objective and corresponding likelihood in Appendix~\ref{appendix:diffusion}. 

\subsubsection{Schedule Alignment} \label{schedule}

In FastDiff, similar to DDPMs, during training we use $T=1000$ discrete time steps. Therefore, when needed to condition on $t$ during sampling, we also need to approximate $T_m$ discrete time indices by aligning the $T_m$-step sampling noise schedule $\hat{\mBeta}$ to the $T$-step training noise schedule $\mBeta$, with $N<<T$. We have attached the detailed algorithms in Appendix~\ref{appendix:Training}. 



\subsection{Training, Noise Scheduling and Sampling}

All illustrated in Algorithm~\ref{alg: training}, we separately parameterize FastDiff by two modules: 1) a iterative refinement model $\theta$ that minimizes a variational bound of the score function, and 2) a noise predictor $\phi$ that optimizes the noise schedule for a tighter evidence lower bound. For inference, we first derive the tighter and more efficient noise schedules $\hat{\beta}$ via an one-shot noise scheduling procedure, which makes FastDiff achieve orders of magnitude faster at sampling. It has been demonstrated~\cite{lam2022bddm} that the noise schedule searched for as few as 1 sample could be robust enough to maintain a high-quality generation among all samples in testing set. Secondly, we map the continuous noise schedules to discrete time indexes $T_m$ using schedule alignment. Finally, FastDiff iteratively refines gaussian noise to generate high-quality samples with computational efficiency. The detailed information on training, noise scheduling and inference procedures has been presented in Appendix~\ref{appendix:schedule_alignment}.

\subsection{FastDiff-TTS} \label{FastDiff_TTS}
Existing text-to-speech methods usually adopt a two-stage pipeline: 1) A text-to-spectrogram generation module (a.k.a. acoustic model) aims to generate prosodic attributes according to variance prediction; 2) A conditional waveform generation module (a.k.a. vocoder) adds the phase information and synthesizes a detailed waveform. To further simplify the text-to-speech synthesis pipeline, we propose a fully end-to-end model FastDiff-TTS, which does not require intermediate features or specialized loss functions. FastDiff-TTS is designed to be a fully differentiable and efficient architecture that directly produces waveforms from contexts (e.g. phonemes) without needing to generate acoustic features (e.g., Mel-spectrograms) explicitly. 

\begin{algorithm}[ht]
    \centering
    \caption{Training refinement network $\theta$}\label{alg: training}
    \begin{algorithmic}[1]
     \STATE \textbf{Input}:  Pre-defined noise schedule $\beta$
    \REPEAT 
    \STATE Sample $\vx_{0} \sim q_{data}$, $\rvepsilon\sim\gN(\vzero,\mI)$, and $t\sim\mathrm{Unif}(\{1,\cdots,T\})$
    \STATE $\vx_t = \alpha_{t} \vx_{0}+ \sqrt{1-\alpha_{t}^2} \rvepsilon$
    \STATE Take gradient descent steps on $\nabla_{\theta}\left\|\rvepsilon-\rvepsilon_{\theta}\left(\vx_t|c, t\right)\right\|_{2}^{2}$ 
    \UNTIL{refinement model $\theta$ converged}
  
    \end{algorithmic}
    \end{algorithm}

    \begin{algorithm}[ht]
        \centering
        \caption{Training noise predictor $\phi$}\label{alg: training_noise}
        \begin{algorithmic}[1]
        \STATE \textbf{Input}: Pre-defined discrete $\beta$, trained refinement network $\theta$, hyperparameter $\tau$.
        \REPEAT 
        \STATE Sample $\vx_{0} \sim q_{data}$, $\rvepsilon\sim\gN(\vzero,\mI)$, and $t\sim\mathrm{Unif}(\{\tau,\cdots,T-\tau\})$
        \STATE $\vx_t = \alpha_{t} \vx_{0}+ \sqrt{1-\alpha_{t}^2} \rvepsilon$
        \STATE $\hat{\beta}_{t}=\min \left\{1-\alpha_{t}^2, 1-\frac{\alpha_{t+\tau}^{2}}{\alpha_{t}^{2}}\right\} \phi(\vx_{t})$
        \STATE Take gradient descent steps on $\nabla_{\phi}\left\{\frac{\delta_{t}^{2}}{2\left(\delta_{t}^{2}-\hat{\beta}_{t}\right)}\left\|\epsilon-\frac{\hat{\beta}_{t}}{\delta_{t}^{2}} \epsilon_{\theta}\left(\boldsymbol{x}_{t}|c, t\right)\right\|_{2}^{2}\right\}$ 
        \UNTIL{noise predictor $\phi$ converged}
        \end{algorithmic}
        \end{algorithm}
\begin{algorithm}[ht]
    \centering
    \caption{Sampling}\label{alg: sampling}
    \begin{algorithmic}[1]
     \STATE \textbf{Input}: Searched $\hat{\beta}$ in noise scheduling process.
    \STATE Compute discrete steps $T_m$ sequences via schedule alignment in Section~\ref{schedule}.
    \STATE Sample $\vx_{T_m} \sim \gN(\vzero,\mI)$
    \FOR{$t=T_m,\cdots,1$}
    \STATE Sample $\vx_{t-1}\sim p_{\theta}(\vx_{t-1}|\vx_t;\hat{\beta})$
    \ENDFOR
    \RETURN $\vx_0$
    \end{algorithmic}
    \end{algorithm}
    
\subsubsection{Architecture}
The architecture design of FastDiff-TTS refers to a convectional non-autoregressive text-to-speech model -- FastSpeech 2~\cite{ren2020fastspeech} as the backbone. The architecture of FastDiff-TTS is illustrated in Figure~\ref{fig:arch}(d). In FastDiff-TTS, the encoder first converts the phoneme embedding sequence into the phoneme hidden sequence. Then, the duration predictor expands the encoder output to match the length of the desired waveform output. Given the aligned sequence, the variance adaptor adds pitch information into the hidden sequence. Note that it is difficult to use the full audio corresponding to the full text sequence for training due to the typically high sampling rate for high-fidelity waveform (i.e., 24,000 samples per second) and the limited GPU memory. Therefore, we sample a small segment to synthesize the waveform before passing to the FastDiff model. Finally, the FastDiff model decodes the adapted hidden sequence into a speech waveform as in the vocoder task.

\subsubsection{Training Loss}
FastDiff-TTS does not require specialized loss functions and adversarial training to improve sample quality as suggested by the previous works~\cite{ren2020fastspeech,donahue2020end,kim2021conditional}. This, to a large extend, simplifies the text-to-speech generation. The final training loss consists of the following terms: 1) a duration prediction loss $L_\text{dur}$: the mean squared error between the predicted and the ground-truth word-level duration in log-scale, 2) a diffusion loss $L_\text{diff}$: the mean squared error between the estimated and gaussian noise, and 3) a pitch reconstruction loss $L_\text{pitch}$: the mean squared error between the predicted and the ground-truth pitch sequences. We empirically found that the pitch reconstruction loss $L_\text{pitch}$ is helpful for handling the one-to-many mapping issue in text-to-speech generation. 

\begin{figure}
    \centering
    \includegraphics[width=0.45\textwidth]{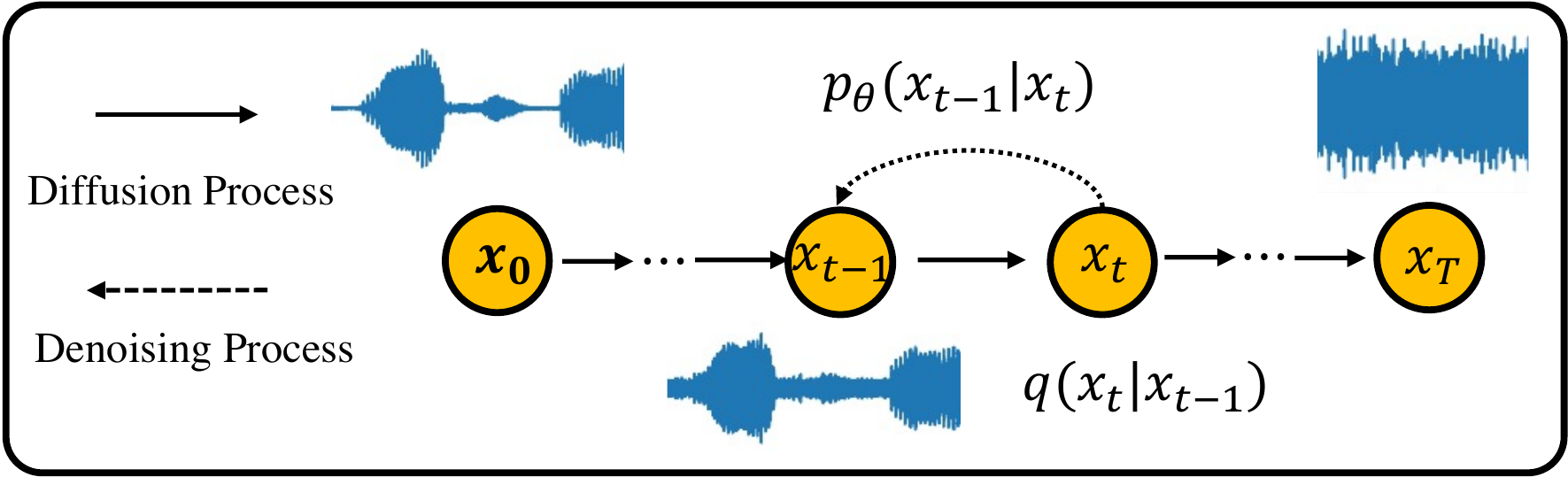}
   \caption{Conditional Diffusion Model for Speech Synthesis} 
    \label{fig:process}
  \end{figure}

\section{Related Works}

Text-to-speech (TTS) systems aim to synthesize raw speech waveforms from given text. In recent years, Neural network based TTS~\cite{ren2020fastspeech,kim2020glow,liu2021diffsinger} has made huge progress and attracted a lot of attention in the machine learning and speech community.

Neural vocoder plays the most important role in the recent success of speech synthesis, which require diverse receptive field patterns to catch audio dependencies: 1) autoregressive model WaveNet~\cite{oord2016wavenet} requires causal convolutions layers and large filters to increase the receptive field while suffering from slow inference speed. 2) Flow-based generative models~\cite{prenger2019waveglow} fully utilize modern parallel computing processors to broaden corresponding receptive fields and speed-up sampling, while they usually achieve a limited sample quality. 3) Generative adversarial networks (GANs)~\cite{jang2021univnet,kong2020hifi} are one of the most dominant deep generative models in audio generation. UnivNet~\cite{jang2021univnet} has demonstrated its success in using local-variable convolution on different waveform intervals, and HIFI-GAN~\cite{kong2020hifi} proposes multi-receptive field fusion (MRF) to model the periodic patterns matters. However, GAN-based models are often difficult to train, collapsing~\cite{creswell2018generative} without carefully selected hyperparameters and regularizers, and showing less sample diversity. 4) Recently proposed diffusion models Diffwave~\cite{kong2020diffwave} and WaveGrad~\cite{chen2020wavegrad} could generate high-quality speech samples, while suffering from a distinct degradation when reducing reverse iterations, making diffusion models difficult to get accelerated. Different from vocoders mentioned above, FastDiff improves the robustness of conditional diffusion model by catching the details of noisy samples at dynamic dependencies, and reduces reverse iterations with predicted noise schedule. The proposed conditional diffuion model allows the high-quality speech synthesis with computational efficiency. 

Another important line of work covers directly waveform generation from text: FastSpeech 2s~\cite{ren2020fastspeech} and VITS~\cite{kim2021conditional} adopt adversarial training process and spectral losses for improving audio quality, while they do not take full advantage of end-to-end training. Recently proposed WaveGrad 2~\cite{chen2021wavegrad} estimates the gradient of the log conditional density of the waveform given a phoneme sequence, but suffers from a large model footprint and slow inference. Unlike all of the aforementioned methods, as highlighted in section~\ref{FastDiff_TTS}, FastDiff-TTS is a fully differentiable and efficient architecture that produces waveforms directly without generating middle features (e.g., spectrograms) explicitly. In additional, our diffuion probabilistic model gets free from hundred or thousands of iterations and enjoy computational efficiency.

\section{Experiments}

\subsection{Setup}

\subsubsection{Dataset}
For a fair and reproducible comparison against other competing methods, we used the benchmark LJSpeech dataset~\cite{LJSpeech}. LJSpeech consists of 13,100 audio clips of 22050 Hz from a Female speaker with about 24 hours in total. To evaluate the generalization ability of our model over unseen speakers in multi-speaker scenarios, we also used the VCTK dataset~\cite{yamagishi2019cstr}, which was downsampled to 22050 Hz to match the sampling rate with the LJSpeech datset. VCTK consists of approximately 44,200 audio clips uttered by 109 native English speakers with various accents. For both datasets, we used 80-band Mel-spectrograms as the condition for the vocoding task. The FFT size, window size, and hop size were, respectively, set to 1024, 1024, and 256. 

\subsubsection{Model Configurations}
FastDiff mainly consists of the refinement model $\theta$ and noise schedule predictor $\phi$. The refinement model $\theta$ comprises three Diffusion-UBlock and DBlock with the upsample or downsample rate of $[8, 8, 4]$, respectively. We adopt a lightweight GALR network effective in separating the added gaussian noise from audio as the noise schedule predictor $\phi$. For end-to-end text-to-speech generation, FastDiff-TTS follows the basic structure in FastSpeech 2~\cite{ren2020fastspeech}, which consists of 4 feed-forward transformer blocks in the phoneme encoder. More details have been attached in the Appendix~\ref{appendix:arch}. 

\subsubsection{Training and Evaluation}
The complete training pipeline has been illustrated in Algorithm~\ref{alg: training}: FastDiff was trained with constant learning rate $lr = 2\times 10^{-4}$. The refinement model $\theta$ and noise predictor $\phi$ were trained for 1M and 10K steps until convergence, respectively. FastDiff-TTS was trained until 500k steps using the AdamW optimizer with $\beta_{1}=0.9,\beta_{2}=0.98,\epsilon=10^{-9}$. Both models were trained on 4 NVIDIA V100 GPUs using random short audio clips of 16,000 samples from each utterance with a batch size of 16 each GPU. More details have been attached in the Appendix~\ref{appendix:Training}.

We crowd-sourced 5-scale MOS tests via Amazon Mechanical Turk to evaluate the audio quality. The MOS scores were recorded with 95\% confidence intervals (CI). Raters listened to the test samples randomly, where they were allowed to evaluate each audio sample once. We further include additional objective evaluation metrics including  STOI and PESQ to test sample quality. To evaluate the sampling speed, we implemented real-time factor (RTF) accessment on a single NVIDIA V100 GPU. In addition, we employed two metrics NDB and JSD to explore the diversity of generated mel-spectrograms. More information about both objective and subjective evaluation has been attached in Appendix~\ref{appendix:evaluation}.

\begin{table*}[]
  \centering
  \small
  \begin{minipage}[t]{0.7\linewidth}
  \centering
  \scalebox{0.9}{
    \begin{tabular}{l|ccc|c|cc}
    \toprule
    \bfseries \multirow{2}{*}{Model} & \multicolumn{3}{c|}{\bf Quality} & \multicolumn{1}{c|}{\bf Speed} & \multicolumn{2}{c}{\bf Diversity} \\
            & MOS ($\uparrow$) & STOI ($\uparrow$) & PESQ ($\uparrow$)  & RTF ($\downarrow$)&NDB ($\downarrow$)&JSD ($\downarrow$) \\
    \midrule
    GT               &  4.52$\pm$0.09   &  /     &  /      & /          & /   & /        \\
    \midrule
    WaveNet (MOL)    &  4.20$\pm$0.06    &  /    &  /     &  85.230    & {33}  & {0.002}\\
    WaveGlow         & 3.89$\pm$0.07     & 0.961 & 3.16   &  0.029    & 66           & 0.014\\
    HIFI-GAN         & 4.08$\pm$0.08     & 0.956 & 3.28   & 0.002     & 72           &  0.010\\
    UnivNet          & 4.13$\pm$0.09     & 0.971 & 3.45   &  0.002    & 68           & 0.013\\
    \midrule
    Diffwave (6 steps)     & 4.18$\pm$0.08   & 0.966 & 3.62   & 0.093       & 72            & 0.007\\
    WaveGrad (50 steps)    & 4.09$\pm$0.07   & 0.911 & 2.70   & 0.390       & 61            & 0.008\\
    \midrule
    FastDiff (4 steps)      &\textbf{4.28$\pm$0.07}& \textbf{0.976} &\textbf{3.71}  &  0.017  & 49 & 0.006 \\
    \bottomrule   
    \end{tabular}}
    \protect\caption{Comparison with other nerual vocoders in terms of quality, synthesis speed and sample diversity. For sampling, we used 50 steps in WaveGrad and 6 steps in DiffWave, respectively}
    \label{table:mos}
  \end{minipage}
  \hspace{0.5cm}
  \begin{minipage}[t]{0.25\linewidth}
  \centering
  \small
    \scalebox{0.9}{
            \begin{tabular}{lcc}
              \toprule
              \bf{Model}                   & MOS \\
              \midrule
              GT                     &  4.37$\pm$0.06  \\
              \midrule
              WaveNet (MOL)          &  4.01$\pm$0.08  \\
              WaveGlow               &  3.66$\pm$0.08  \\
              HIFI-GAN               &  3.74$\pm$0.06   \\
              UnivNet                &  3.85$\pm$0.07   \\
              \midrule
              Diffwave (6 steps)               &  3.90$\pm$0.07   \\
              WaveGrad (50 steps)               &  3.72$\pm$0.06    \\
              \midrule
              FastDiff (4 steps)               & \textbf{4.10$\pm$0.06}  \\
              \bottomrule
              \end{tabular}}
              \vspace{1mm}
              \protect\caption{Comparison with other neural vocoders of synthesized utterances for unseen speakers.}
              \label{table:mos4}
          \end{minipage}
    \end{table*}
    
    \begin{table*}[ht]
        \centering
        \small
        \begin{minipage}[t]{.5\linewidth}
        \centering
        \scalebox{0.9}{
        \begin{tabular}{lccccc}
        \toprule
        \bfseries Model   & MOS ($\uparrow$)   & STOI($\uparrow$) & PESQ ($\uparrow$) & RTF ($\downarrow$) \\
        \midrule
        GT                   &  4.52$\pm$0.09           & /  & /   & /  \\
        \midrule
        w/o Time-aware LVC                 & 4.08$\pm$0.05         & 0.971         & 3.45   & 0.081  \\
        w/o Noise Predictor                  & 4.10$\pm$0.06         & 0.968         & 3.50   & 0.033    \\
        \midrule
        Continuous, 4 steps     &  4.09$\pm$0.08        & 0.970         & 3.37   &\textbf{0.015} \\
        Continuous, 1000 steps &  4.14$\pm$0.07        & 0.980         & 3.64   &  3.80    \\
        \midrule
        Discrete, 4 steps      & 4.28$\pm$0.07          & 0.976         & 3.71    & 0.017 \\
        Discrete, 1000 steps   & \textbf{4.36$\pm$0.08} & \textbf{0.989} & \textbf{3.86}  & 4.70 \\
        \bottomrule
      \end{tabular}}
      \protect\caption{Ablation study results. Comparison of the effect of each component in terms of quality and synthesis speed.}
      \label{table:mos2}
    \end{minipage}
      \hspace{0.6cm}
      \begin{minipage}[t]{0.3\linewidth}
      \centering
      \small
      \scalebox{0.95}{
      \begin{tabular}{lcc}
        \toprule
        \bf{Model}                   & MOS \\
        \midrule
        GT                     &  4.52$\pm$0.09   \\
        GT(voc.)               &  4.28$\pm$0.07   \\
        Cascaded               &  4.13$\pm$0.07   \\
        \midrule
        FastSpeech 2s          &  3.94$\pm$0.06  \\
        WaveGrad 2             &  3.68$\pm$0.09   \\
        \midrule
        FastDiff-TTS           &  \textbf{4.03$\pm$0.09}  \\
        \bottomrule
        \end{tabular}}
        \vspace{1mm}
        \protect\caption{Comparison with other text-to-speech models in terms of quality.}
        \label{table:mos3}
      \end{minipage}
    \vspace{-2mm}
      \end{table*}

\subsection{Comparsion with other models}
We compared our FastDiff in audio quality, diversity, and sampling speed with competing models, including 1) WaveNet~\cite{oord2016wavenet}, the autoregressive generative model for raw audio. 2) WaveGlow~\cite{prenger2019waveglow}, non-autoregressive flow-based model. 3) HIFI-GAN V1~\cite{kong2020hifi} and UnivNet~\cite{jang2021univnet}, the most dominant and popular GAN-based models. 4) Diffwave~\cite{kong2020diffwave} and WaveGrad~\cite{chen2020wavegrad}, recently proposed diffusion probabilistic models which achieve state-of-the-art in speech synthesis. For easy comparison, the results are compiled and presented in Table~\ref{table:mos}, and we have the following observations: 

In terms of audio quality, FastDiff achieved the highest MOS with a gap of $0.24$ compared to the ground truth audio, and it matched the performance of the autoregressive WaveNet baseline and outperformed the non-autoregressive baselines. For objective evaluation, FastDiff also demonstrated a large improvement in PESQ and STOI. For inference speed, with the efficient noise schedules searched by noise predictor, FastDiff could generate high-quality speech samples within as few as $4$ reverse steps, significantly reducing the inference time compared with competing diffusion architectures. To the best of our knowledge, FastDiff makes diffusion models for the first time applicable to interactive, high-quality real-world speech synthesis at a low computational cost. In terms of sample diversity, we can see that FastDiff still witnessed a gap from autoregressive WaveNet, but it achieve a higher variety for generated speeches than non-autoregressive baselines. More detailed evaluation on sample diversity has been attached in Appendix~\ref{appendix:diversity}.

\subsection{Ablation study}\label{ablation}
We conducted ablation studies to demonstrate the effectiveness of several designs in FastDiff, including the time-aware location variable convolution and noise predictor in neural vocoding. The results of both subjective and objective evaluations have been presented in Table~\ref{table:mos2}, and we have the following observations: 1) Replacing time-aware location-variable convolution by traditional convolutional operations causes a distinct degradation in sampling speed and perceptual quality. 2) Using grid search instead of the noise predictor to search schedules had witnessed the decreased audio quality, demonstrating that the noise schedule prediction process provides more efficient reverse sampling without sacrificing quality.

Further, we compare two variants of FastDiff to test the modality differences of diffusion condition (i.e., continuous noise-level or discrete time-step). Note that the former model does not require the schedule alignment process mentioned in Section~\ref{schedule}. We empirically find that the FastDiff model conditioned on discrete time steps could synthesize samples with higher quality, demonstrating that learning proposed FastDiff with discrete diffusion times could be a better choice. More information on the variant of FastDiff extended to continuous noise schedules has been attached in Appendix~\ref{appendix:extension}

\subsection{Generalization to unseen speakers}\label{Generalization}

We used $50$ randomly selected utterances of $5$ unseen speakers in the VCTK dataset that were excluded from the training set for the MOS test. Table~\ref{table:mos4} shows the experimental results for the mel-spectrogram inversion of the unseen speakers: In summary, we noticed that FastDiff achieved state-of-the-art in terms of audio quality for out-of-domain generalization, indicating that FastDiff could universally generate high-fidelity audio from entirely new (unseen) speakers outside the train set. 

\subsection{End-to-End Text-to-Speech}

To demonstrate the robustness of the proposed model in end-to-end text-to-speech synthesis, we compare FastDiff-TTS with other neural TTS systems, including 1) GT, the ground truth audio; 2) GT (voc.), where we first convert the ground truth audio into mel-spectrograms, and then convert the mel-spectrograms back to audio using FastDiff; 3) PortaSpeech~\cite{ren2021portaspeech} + FastDiff: vocoder cascaded with mel-spectrogram generation using the most popular non-autoregressive TTS models; 4) FastSpeech 2s~\cite{ren2020fastspeech}: the extension of FastSpeech 2 to fully end-to-end text-to-waveform generation with multi-task learning; 5) WaveGrad 2~\cite{chen2021wavegrad}: a diffusion probabilistic model to generate waveforms via gradient estimation. The results are shown in Table~\ref{table:mos3}: FastDiff-TTS could surpass competing end-to-end speech synthesis models and match the voice quality of the state-of-the-art cascaded TTS systems, demonstrating that FastDiff-TTS is efficient in simplifying the overall text-to-speech synthesis pipeline.
\section{Conclusion}

In this work, we proposed FastDiff, a fast conditional diffusion model for high-quality speech synthesis. FastDiff employed a stack of time-aware location-variable convolutions with diverse receptive field patterns to model long-term time dependencies with adaptive conditions. A noise predictor was further adopted to derive tighter schedules for reducing reverse iterations without distinct quality degradation. The extension model FastDiff-TTS discarded intermediate features (e.g., spectrograms) and simplified the end-to-end text-to-waveform syntheses pipeline. Experimental results demonstrated that our proposed model outperformed the best publicly available models in terms of synthesis quality, even comparable to the human level.
Moreover, FastDiff showed a significant improvement in synthesis speed, which required as few as $4$ iterations to generate high-quality samples. To the best of our knowledge, FastDiff made diffusion models for the first time applicable to interactive, real-world speech generation with a low computational cost. In addition, FastDiff performed strong robustness and enjoyed high-quality synthesis in out-of-domain generalization to unseen speakers. We will release our code and pre-trained models in the future, and we envisage that our work could serve as a basis for future speech synthesis studies.

\clearpage
\bibliographystyle{named}
\bibliography{ijcai22}

\appendix
\clearpage

\section{Diffusion Probabilistic models} \label{appendix:diffusion}
Given i.i.d. samples $\{\mathbf{x}_{0} \in \mathbb{R}^{D}\}$ from an unknown data distribution $p_{data}(\mathbf{x}_{0})$. In this section, we introduce the theory of diffusion probabilistic model~\cite{ho2020denoising,lam2022bddm,song2020denoising,song2020score}. First, we present diffusion and reverse process given by denoising diffusion probabilistic models (DDPMs), which could be used to learn a model distribution $p_{\theta}(\mathbf{x}_{0})$ that approximates $p_{data}(\mathbf{x}_{0})$. Secondly, we introduce the recently proposed bilateral denoising diffusion models (BDDMs) and its tighter evidence lower bound (ELBO) for acceleration. 

\paragraph{Diffusion process}  Similar as previous work~\cite{ho2020denoising,lam2022bddm,song2020denoising}, we define the data distribution as $q(\mathbf{x}_{0})$. The diffusion process is defined by a fixed Markov chain from data $x_0$ to the latent variable $x_T$:
\begin{align}
q(\vx_{1},\cdots,\vx_T|x_0) = \prod_{t=1}^T q(\vx_t|\vx_{t-1}),
\quad\ \ 
\end{align}

For a small positive constant $\beta_t$, a small Gaussian noise is added from $x_{t}$ to the distribution of $x_{t-1}$ under the function of $q(x_t|x_{t-1})$.

The whole process gradually converts data $x_0$ to whitened latents $x_T$ according to the fixed noise schedule $\beta_1,\cdots,\beta_T$.
\begin{align}
q(\vx_t|\vx_{t-1}) := \gN(\vx_t;\sqrt{1-\beta_t}\vx_{t-1},\beta_t I)
\end{align}

Efficient training is optimizing a random term of $t$ with stochastic gradient descent: 

\begin{align}
    \label{eq: score_loss}
    \gL_{\theta} = \left\lVert \beps_\theta\left(\alpha_t\rvx_{0}+\sqrt{1-\alpha_t^2}\beps\right)-\beps\right\rVert_2^2, \beps\sim\gN(\vzero, \mI)
\end{align}

\paragraph{Reverse process}  Unlike the diffusion process, reverse process is to recover samples from Gaussian noises. The reverse process is a Markov chain from $x_T$ to $x_0$ parameterized by shared $\theta$:
\begin{align}
p_{\theta}(\vx_0,\cdots,\vx_{T-1}|x_T)=\prod_{t=1}^T p_{\theta}(\vx_{t-1}|\vx_t),
\end{align}

where each iteration eliminate the Gaussian noise added in the diffusion process:
\begin{align}
    p(\vx_t|\vx_{t-1}) := \mathcal{N}(\vx_{t-1};\mu_{\theta}(\vx_t,t), \sigma_{\theta}(\vx_t,t)^2I)
\end{align}

\paragraph{Acceration} 
Recently, Bilateral denoising diffusion models (BDDMs)~\cite{lam2022bddm} demonstrates its tighter evidence lower bound (ELBO) for noise schedule prediction. Given a leaned diffusion network $\theta$, a scheduling network $\phi$ could be applied in reducing the gap between the proposed surrogate objective. To be more specific, instead of using the fixed one in diffusion process, a much more efficient N-step noise schedule (i.e., $\hat{\beta}$) could be derived by the well-leaned noise scheduling network $\phi$. The noise schedule could be applied in reverse process, making it possible to explicitly trade-off between inference computation and output quality in one model. 

For learning the noise schedule predictor $\phi$, we apply the loss function as a KL divergence term between the forward and the reverse distribution:
\begin{equation} \label{eq: noise_loss}
    \resizebox{.99\linewidth}{!}{$
    \gL_{\phi}= \frac{1}{2(1-\beta_{t}-\alpha_{t}^{2})}\|\sqrt{1-\alpha_{t}^{2}} \beps_{t}-\frac{\beta_{t}}{\sqrt{1-\alpha_{t}^{2}}} \beps_{\theta}(\vx_{t}, \alpha_{t})\|_{2}^{2} +C_{t}
    $}
\end{equation}
where $C_{t} =\frac{1}{4} \log \frac{1-\alpha_{t}^{2}}{\beta_{t}}+\frac{D}{2}(\frac{\beta_{t}}{1-\alpha_{t}^{2}}-1)$ is a constant that can be ignored during training.

\section{Model Architectures} \label{appendix:arch}
\subsection{FastDiff} 
\label{appendix:arch_refime} \label{appendix:noise_predictor}

\begin{figure*}
    \vspace{-8mm}
    \centering
    \includegraphics[width=0.6\textwidth,trim={1.0cm 0cm 1.0cm 0cm}]{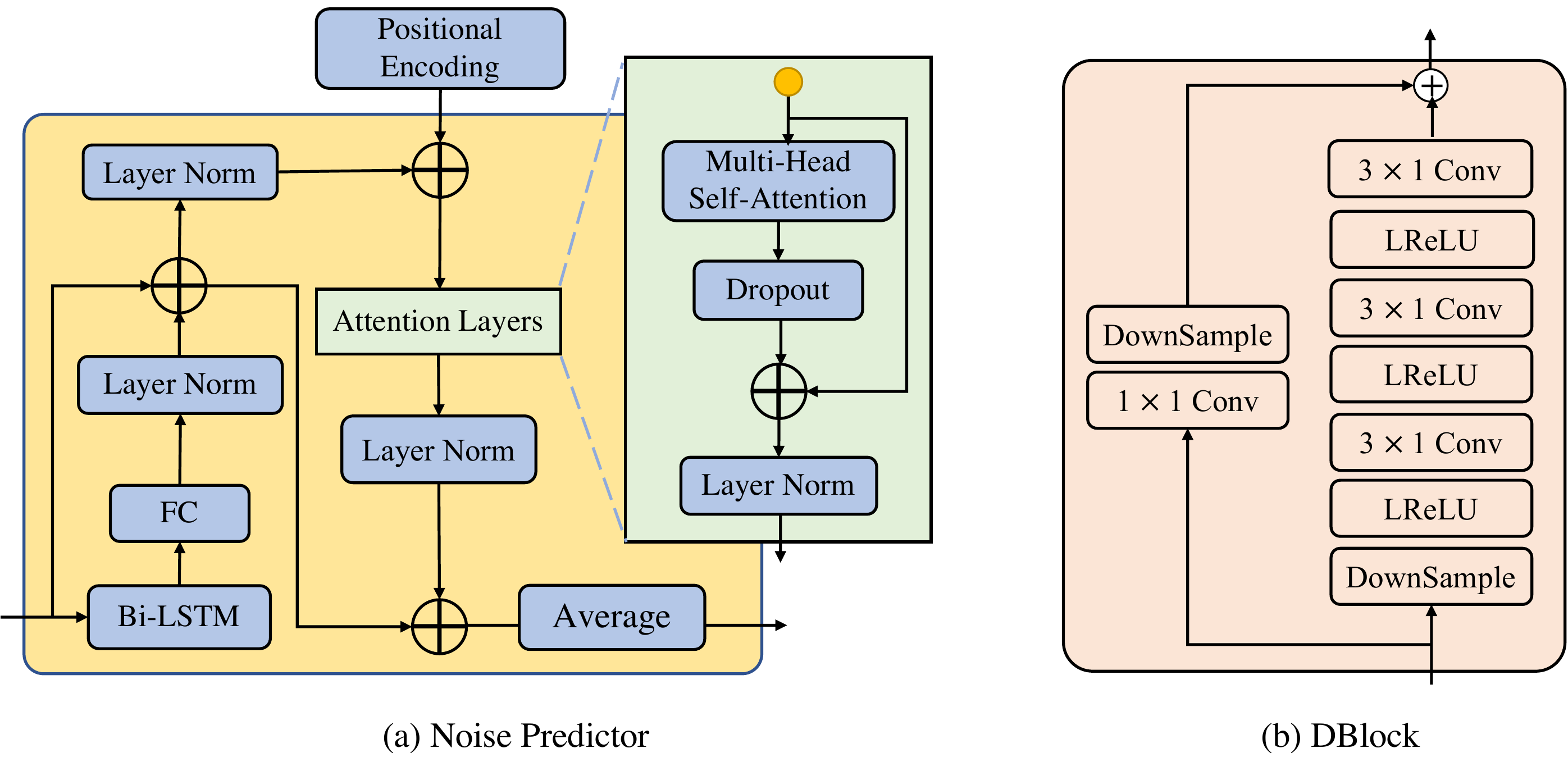}
    \vspace{-3mm}
   \caption{The details of network architectures. Left: The GALR-block based noise predictor $\phi$. Right: DBlock in FastDiff $\theta$} 
    \label{fig:noise_predictor}
  \end{figure*}

As illustrated in Table \ref{tab:hyperparameters}, we list the hyper-parameters of FastDiff. We further visualize the detailed architectures of the noise predictor and DBlock in the refinement model in Figure~\ref{fig:noise_predictor}.

\begin{table}[h]
\centering
\begin{minipage}[t]{\linewidth}
\centering
\begin{tabular}{l|l}
\hline
\textbf{Hyperparameter}               &  \textbf{FastDiff}  \\
\toprule
\textbf{Refinement Model $\theta$}    &               \\
DBlock Hidden Channels                 &     32     \\ 
DBlock Downsample Ratios               &  [4, 8, 8]           \\ 
Diffusion UBlock Hidden Channels      &    32          \\
Diffusion UBlock Upsample Ratios       &    [8, 8, 4]    \\
Time-aware LVC layers Each Block       &      4         \\
Time-aware LVC layers Kernel Size      &      256       \\
Diffusion Kernel Predictor Hidden Channels   &       64   \\  
Diffusion Kernel Predictor Kernel Size  &      3       \\
Diffusion Embedding Input Channels      &      128     \\
Diffusion Embedding Output Channels     &      512     \\
Use Weight Norm                         &      True          \\
\midrule
Total Number of Parameters             &           13M	         \\ 
\midrule
\textbf{Noise Predictor $\phi$}      &               \\
Window Length                       &  8 Samples             \\
Segment Size                        &  64             \\
Number of GALR Blocks               &  2             \\
GALR Blocks Hidden Channels         &  128             \\
\midrule
Total Number of Parameters          &   0.5M         \\ 
\bottomrule
\end{tabular}
\caption{Architecture hyperparameters of FastDiff.}
\label{tab:hyperparameters}
\end{minipage}
\end{table}



\subsection{FastDiff-TTS} \label{appendix:fastDiff_tts}
In this section, we list the model hyper-parameters of FastDiff-TTS in Table \ref{tab:hyperparameters}.

\begin{table}[H]
    \begin{minipage}[t]{\linewidth}
        \centering
        \begin{tabular}{l|l}
        \hline
        \textbf{Hyperparameter}               &  \textbf{FastDiff-TTS}  \\
        \toprule
        Phoneme Embedding Dimension             &  256                   \\ 
        Pre-net Layers                          &      3                         \\ 
        Pre-net Hidden                        &     256                          \\ 
        Encoder Layers                         &      4                       \\ 
        Encoder Hidden                         &      256                         \\
        Encoder Conv1D Kernel                  &       9                         \\ 
        Encoder Conv1D Filter Size             &      1024                         \\
        Encoder Attention Heads                &      2                       \\
        Encoder/Decoder Dropout                &      0.1                     \\
        Variance Predictor Conv1D Kernel       &  3                          \\ 
        Variance Predictor Conv1D Filter Size  &  256                      \\ 
        Variance Predictor Dropout              &   0.5                 \\ 
        FastDiff wave decoder               &   Follow Table~\ref{tab:hyperparameters}              \\ 
        \midrule
        Total Number of Parameters            &           40M	         \\
        \bottomrule
        \end{tabular}
        \caption{Architecture hyperparameters of FastDiff-TTS.}
        \label{tab:hyperparameters_tts}
        \end{minipage}
\end{table}

\section{Training, Noise scheduling and Inference details}  \label{appendix:Training}

\subsection{Diffusion hyperparameters} \label{appendix:noise_schedule}
We list the diffusion hyper-parameters of FastDiff and FastDiff/FastDiff-TTS in Table \ref{tab:apx_diffusionhyperparameters}.

\begin{table}[h]
    \small
    \centering
    \begin{tabular}{l}
    \hline
    \textbf{Diffusion Hyperparameter}       \\
    \toprule 
    \textbf{Noise Scheduling}                \\
    $\tau = 200, \hat{\alpha_{t}} = 0.54, \hat{\beta_{t}} = 0.70, N = 4$                                            \\  
    \midrule
    \textbf{Training and Sampling}                            \\
    \textbf{Pre-defined (T = $1000$)}: \\ $\beta= \text{Linear} (1\times 10^{-4}, 0.005, 1000)$       \\
    \textbf{Grid Search ($T_m = 4$) derived}:  \\ $ \hat{\beta} = [3.6701e^{-7}, 1.7032e^{-5}, 7.908e^{-4}, 7.6146e^{-1}]$      \\
    \textbf{Noise Predictor ($T_m = 4$) derived}: \\ $ \hat{\beta} = [3.2176e^{-4}, 2.5743e^{-3}, 2.5376e^{-2}, 7.0414e^{-1}]$      \\
    \bottomrule
    \end{tabular}
    \caption{Diffusion hyperparameters of FastDiff and FastDiff-TTS.}
    \label{tab:apx_diffusionhyperparameters}
    \end{table}

\subsection{Noise Scheduling}  \label{appendix:algorithm}
Our noise scheduling algorithm mainly follows the bilateral denoising diffusion models~\cite{lam2022bddm}:
\begin{algorithm}[H]
    \centering
    \caption{Noise scheduling process}\label{alg: noise scheduling}
    \begin{algorithmic}[1]
    \STATE  \textbf{Input}: Pre-defined discrete $\beta$, trained refinement network $\theta$, hyperparameter $N, \hat{\alpha_{t}}, \hat{\beta_{t}}$.
    \FOR{$t=N,\cdots,2$}
    \STATE Sample $\vx_{t-1}\sim p_{\theta}(x_{t-1}|x_t)$
    \STATE  $\hat{\alpha}_{t-1}=\frac{\hat{\alpha}_{t}}{\sqrt{1-\hat{\beta}_{t}}}$
    \STATE  $\hat{\beta}_{t-1}=\min \left\{1-\hat{\alpha}_{t-1}^{2}, \hat{\beta}_{t}\right\} \phi\left(\hat{\boldsymbol{x}}_{t-1}\right)$
    \IF{$\hat{\beta}_{t-1}<\beta_{1}$}
    \RETURN $\hat{\beta}_{t}, \ldots, \hat{\beta}_{N}$
    \ENDIF
    \ENDFOR
    \RETURN $\hat{\beta}_{t}, \ldots, \hat{\beta}_{N}$
    \vspace{0.09em}
    \end{algorithmic}
\end{algorithm}

\subsection{Schedule Alignment}  \label{appendix:schedule_alignment}
Here we search and interpolate $\alpha_{s}$ between two training noise constants $l_{t}$ and $l_{t+1}$, enforcing $\alpha_{s}$ to get closed to $l_{t}$. In the end, we gain the well-mapped diffusion step $t_m$:

Firstly we compute the corresponding constants respective to diffusion and reverse process:
\begin{align}
    l_{t}=\prod_{i=1}^{t} \sqrt{1-\beta_{i}}, \quad \alpha_{s}=\prod_{i=1}^{s} \sqrt{1-\hat{\beta}_{i}}
\end{align}

Here we search and interpolate $\alpha_{s}$ between two training noise constants $l_{t}$ and $l_{t+1}$, enforcing $\alpha_{s}$ to get closed to $l_{t}$. In the end, we gain the well-mapped diffusion step $t_m$:

\begin{equation}\label{eq: aligned step}
    t_m = t + \frac{l_t-\alpha_s}{l_t-l_{t+1}} \quad \text{ if } \alpha_s\in[\ l_{t+1}, l_t\ ].
\end{equation}

Where integer $t$ represents a single pre-defined diffusion step, and $s$ presents a single step of noise schedule obtained through the scheduling process. Given these two schedules mentioned above, we could conduct schedule alignment and derive the floating-point $t_m$ for much more efficient reverse sampling.

\begin{figure*}
    \vspace{-2mm}
    \centering
    \includegraphics[width=0.8\textwidth]{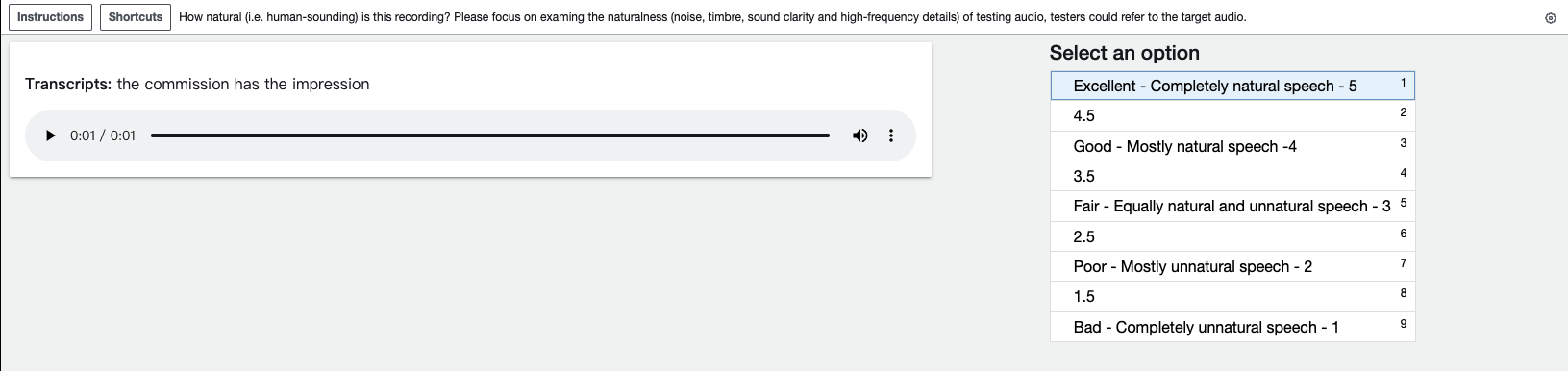}
    \vspace{-1mm}
   \caption{Screenshot of MOS testing.} 
    \label{fig:MOS}
  \end{figure*}
\section{Evaluation Matrix} \label{appendix:evaluation}


\subsection{PESQ and STOI} Perceptual evaluation of speech quality (PESQ)~\cite{rix2001perceptual} and The short-time objective intelligibility (STOI)~\cite{taal2010short} assesses the denoising quality for speech enhancement.

\subsection{NDB and JSD}
Number of Statistically-Different Bins (NDB) and Jensen-Shannon divergence (JSD). They measure diversity by 1) clustering the training data into several clusters, and 2) measuring how well the generated samples fit into those clusters.

\subsection{Details in MOS Evaluation}

All our Mean Opinion Score (MOS) tests are crowd-sourced and conducted by native speakers. The scoring criteria has been included in Table~\ref{matrix:naturalness} for completeness. The samples are presented and rated one at a time by the testers, each tester is asked to evaluate the subjective naturalness of a sentence on a 1-5 Likert scale. The screenshots of instructions for testers are shown in Figure~\ref{fig:MOS}. We paid \$8 to participants hourly and totally spent about \$750 on participant compensation.

\begin{table}[ht]
  \vspace{-2mm}
 \centering
    \small
  \begin{tabular}{ccc}

  \toprule
  Rating & Naturalness & Definition                           \\
  \midrule
  1      & Bad        &  Very annoying and objectionable dist. \\
  2      & Poor       &  Annoying but not objectionable dist. \\
  3      & Fair       &  Perceptible and slightly annoying dist\\
  4      & Good       & Just perceptible but not annoying dist. \\
  5      & Excellent  & Imperceptible distortions\\
  \bottomrule
  \end{tabular}
  \caption{Ratings that have been used in evaluation of speech naturalness of synthetic and ground truth samples.}
  \label{matrix:naturalness}
  \vspace{-4mm}
  \end{table}

\section{Extension to Continuous Condition} \label{appendix:extension}

Our ablation study extends FastDiff to be conditioned on continuous noise levels and compares it to the basic model with the discrete condition. To be more specific, the FastDiff model conditioned on continuous noise levels does not require an additional schedule alignment process, which has a separated training and sampling procedure: 

\begin{algorithm}[H]
    \centering
    \caption{Training refinement network $\theta$ (Continuous Condition)}
    \begin{algorithmic}[1]
    \vspace{-0.1cm}
    \STATE \textbf{Input}: Pre-defined noise schedule $l$
    \REPEAT 
    \STATE Sample $\vx_{0} \sim q_{data}$, $\epsilon\sim\gN(0,I)$, and \quad $t\sim\mathrm{Unif}(\{1,\cdots,T\})$
    \STATE $\alpha_s \sim \operatorname{Uniform}\left(\alpha_{t-1}, \alpha_{t}\right)$
    \STATE $\vx_s = \alpha_{s} \vx_{0}+ \delta_s \epsilon$
    \STATE Take gradient descent steps on $\nabla_{\theta}\left\|\epsilon-\epsilon_{\theta}\left(\vx_s, \alpha_s\right)\right\|_{2}^{2}$ 
    \UNTIL{iterative refinement model $\theta$ converged}
    \vspace{0.09em}
    \end{algorithmic}
    \end{algorithm}

    \begin{algorithm}[H]
        \centering
        \caption{Sampling}
        \begin{algorithmic}[1]
        \STATE \textbf{Input}: Pre-defined $\beta, T$ and $\hat{\beta}$ derived in noise scheduling process.
        \FOR{$t=T,\cdots,1$}
        \STATE Sample $\vx_{t-1}\sim p_{\theta}(\vx_{t-1}|\vx_t)$
        \ENDFOR
        \RETURN $\vx_0$
        \end{algorithmic}
        \end{algorithm}
        \vspace{-2mm}

\section{Sample Diversity} \label{appendix:diversity}
Previous works~\cite{dhariwal2021diffusion,xiao2021tackling} in the image generation task has demonstrated that diffusion probabilistic model outperforms GAN in sample diversity, while the comparison in the speech domain is relatively overlooked. Similarly, we can intuitively infer that diffusion probabilistic models are good at generating high-fidelity diverse speech samples. To verify our hypothesis, we employed two metrics NDB and JSD to explore the diversity of generated mel-spectrograms. As shown in Table~\ref{table:mos}, we can see that diffusion probabilistic model achieve a higher JSD and matching NDB score for generated speeches compare to GAN-based model, which is expected for the following reasons:

1) It is well-known that the mode collapse problem~\cite{creswell2018generative} appears in the dominated GAN-based generative models, which leads to very similar output samples from a single or few modes of the distribution, especially in the strongly conditional generation task.  
2) In contrast, diffusion probabilistic model is meant to reduce mode collapse compared to one-shot generation. It breaks the generation process into several conditional denoising diffusion steps in which each step is relatively simple to model. Thus, we expect our model to exhibit better training stability and mode coverage.

\end{document}